\documentclass[sigconf,natbib=false]{acmart}

\usepackage[backend=biber,maxnames=2]{biblatex} 
\AtBeginDocument{%
  }

\setcopyright{acmlicensed}
\copyrightyear{2018}
\acmYear{2018}
\acmDOI{XXXXXXX.XXXXXXX}
\acmConference[EASE 2025]{The 29th International Conference on Evaluation and Assessment in Software Engineering}{17–20 June, 2025}{Istanbul, Türkiye}
\acmISBN{978-1-4503-XXXX-X/2018/06}

\usepackage{amsmath,amssymb,amsfonts}
\usepackage{graphicx}
\usepackage{textcomp}
\usepackage{xcolor}
\usepackage{multirow}
\usepackage{colortbl}
\usepackage{url}
\usepackage{threeparttable}
\usepackage{array}
\usepackage{float}
\usepackage{titlesec}
\usepackage{fancyvrb}
\usepackage{listings} 
\usepackage{booktabs}
\usepackage{caption}
\usepackage{subcaption}
\usepackage{lineno}
\usepackage{algorithm}

\usepackage{algpseudocode}
\usepackage{amsmath}
\usepackage{tikz}
\usepackage{amsmath}
\usepackage{placeins}

\usepackage{titlesec}
\titlespacing{\section}{0pt}{5pt}{3pt} 
\titlespacing{\subsection}{0pt}{4pt}{2pt} 
\titlespacing{\subsubsection}{0pt}{3pt}{1pt} 

\usepackage{etoolbox}
\AtBeginBibliography{%
  \footnotesize
  \setlength{\bibitemsep}{2pt}
  \setlength{\bibhang}{0.5em}
}
\begin{document}

\title{NRevisit: A Cognitive Behavioral Metric for Code Understandability Assessment}

\author{Gao Hao}
\authornote{The author is a dual-degree student at Macao Polytechnic University and University of Coimbra.}
\affiliation{%
  \institution{Faculty of Applied Sciences, Macao Polytechnic University}
  \city{Macao}
  \country{China}
}
\email{p2109002@mpu.edu.mo}

\author{Haytham Hijazi}
\affiliation{%
  \institution{CISUC, University of Coimbra}
  \city{Coimbra}
  \country{Portugal}}
\email{haytham@dei.uc.pt}

\author{Júlio Medeiros}
\affiliation{%
  \institution{CISUC, University of Coimbra}
  \city{Coimbra}
  \country{Portugal}}
\email{juliomedeiros@dei.uc.pt}

\author{João Durães}
\affiliation{%
\institution{Polytechnic University of Coimbra}
  \city{Coimbra}
  \country{Portugal}}
\email{jduraes@isec.pt}

\author{Chan Tong Lam}
\affiliation{%
  \institution{Faculty of Applied Sciences, Macao Polytechnic University}
  \city{Macao}
  \country{China}
 }
\email{ctlam@mpu.edu.mo}  

\author{Paulo de Carvalho}
\affiliation{%
\institution{CISUC, University of Coimbra}
  \city{Coimbra}
  \country{Portugal}}
\email{carvalho@dei.uc.pt}

\author{Henrique Madeira}
\affiliation{%
\institution{CISUC, University of Coimbra}
  \city{Coimbra}
  \country{Portugal}
  }
\email{henrique@dei.uc.pt}

\renewcommand{\shortauthors}{Gao Hao et al.}

\begin{abstract}
Measuring code understandability is both highly relevant and exceptionally challenging. This paper proposes a dynamic code understandability assessment method, which estimates a personalized code understandability score from the perspective of the specific programmer handling the code. The method consists of dynamically dividing the code unit under development or review in code regions (invisible to the programmer) and using the number of revisits (NRevisit) to each region as the primary feature for estimating the code understandability score. This approach removes the uncertainty related to the concept of a "typical programmer" assumed by static software code complexity metrics and can be easily implemented using a simple, low-cost, and non-intrusive desktop eye tracker or even a standard computer camera. This metric was evaluated using cognitive load measured through electroencephalography (EEG) in a controlled experiment with 35 programmers. Results show a very high correlation ranging from $r_s = 0.9067$ to $r_s = 0.9860$ (with p nearly 0) between the scores obtained with different alternatives of NRevisit and the ground truth represented by the EEG measurements of programmers' cognitive load, demonstrating the effectiveness of our approach in reflecting the cognitive effort required for code comprehension.  The paper also discusses possible practical applications of NRevisit, including its use in the context of AI-generated code, which is already widely used today.
\end{abstract}

\ccsdesc[500]{Software and its engineering~Software maintenance tools}

\keywords{Code complexity metrics, Program comprehension, Cognitive load, Software quality, EEG}


\maketitle
\section{Introduction}
The complexity of the code poses significant challenges in the testing, comprehension and maintenance of software. As complexity increases, it becomes harder to understand and test the code, leading to greater difficulty in maintaining it \cite{van1994general}. This increase in complexity is strongly associated with a higher fault density \cite{chen2019complexity}\cite{omri2018static}, and code complexity is a foundational element in bug prediction techniques
\cite{pachouly2022systematic}\cite{thota2020survey}, playing a critical role in ensuring software quality and development efficiency.

Accurately characterizing code complexity is crucial for determining the appropriateness of code refactoring, ensuring adequate testing (including the selection of the best testing priorities), deciding the best component granularity at the low-level view of the software architecture, keeping code maintainability under control, estimating defect density in software packages, among other classic utilizations of code complexity metrics.

Static code complexity metrics such as Cyclomatic Complexity (V(g)) \cite{mccabe1976complexity} or Cognitive Complexity from SonarSource tools \cite{campbell2018cognitive} are powerful tools because they provide a very easy way to characterize code complexity.
Metrics can be implemented as standalone tools or integrated with IDEs (Integrated Development Environment). All the metric tools operate in a similar way: receive/scan the source code
as input and provide a code complexity score for such code.

However, the reality is that the score provided by static code complexity metrics is always a crude approximation of the real difficulty a specific programmer may feel in understanding the code. Even a potentially perfect static code complexity metric (which does not exist) may always fail for programmers who do not represent the typical programmer, such as novice or top experts. Not to mention the fact that the level of comprehension of a given code depends on other individual factors such as fatigue, distractions, stress and the number and intensity of the interruptions the programmer may have during the code comprehension task in real software development scenarios.

The rapid adoption of LLM-based tools, such as ChatGPT and Copilot, is increasing the use of machine-generated code in software development, introducing additional challenges in code comprehension. Programmers must evaluate and adjust the machine-generated  code to ensure suitability, correctness and readability. Without fully understanding it, they cannot guarantee its quality and correctness. Relying on static code complexity metrics to estimate the difficulty programmers face with machine-generated code is questionable, highlighting the need for new approaches to assess code understandability dynamically (i.e., while the developer is reading and developing the code), such as the one proposed in this paper.

Studies on cognitive complexity and code understandability have shown that structural code complexity alone (as reported by most of the existing code complexity metrics) does not accurately reflect the actual challenges faced by developers during code comprehension \cite{hao2023accuracy}\cite{munoz2020empirical}\cite{Siegmund2021CodeComprehension}. Empirical works using biometric and neuroscience techniques indicate that traditional code complexity metrics often fail in reporting the cognitive difficulties programmers experience in understanding code, highlighting the need for alternative approaches that integrate real-time cognitive engagement
\cite{couceiro2019biofeedback}\cite{medeiros2021can}\cite{peitek2021program}. 

Eye-tracking and EEG-based cognitive load estimation have emerged as promising alternatives, offering a human-centered perspective on code understandability \cite{abbad2022estimating}\cite{aljehane2023studying}\cite{mathur2021dynamic}. However, while EEG directly measures cognitive load, its high intrusiveness limits real-world applicability. Assessing developers’ coding behavior using eye tracking has been the subject of a large body of research works \cite{obaidellah2018survey}, but despite the low-cost and the non-intrusion caused by desktop eye trackers, the reality is that the software industry does not ordinarily integrate eye tracking in IDEs to assess developers cognitive load. A possible reason for the lack of practical solutions is the high complexity of deriving simple metrics from developers' coding behavior.

This paper proposes \textbf{NRevisit}, a practical cognitive-behavioral metric that dynamically assesses developers' cognitive load based on gaze revisit patterns in code regions of the code under development. Unlike static complexity metrics, NRevisit provides a real-time, programmer specific measure of the perceived difficulty in understanding a given code snippet. 
 The intuition is that the number of times the programmer changes the reading focus among the different code regions of the program under analysis (e.g., to confirm the correct understanding of a given code snippet/region or due to meticulous reading) serves as a strong indicator of the cognitive effort required to understand the program. 

To evaluate its effectiveness, we compare two variants of NRevisit metrics: \textbf{C NRevisit}, which targets sustained attention by counting the number of distinct fixations within a code region, and \textbf{CL NRevisit}, which accounts for context-switching by counting only complete exit-return cycles, where the gaze moves to another region before returning. By distinguishing sustained attention (C NRevisit) from context-switching (CL NRevisit), our approach offers deeper insights into code comprehension and cognitive workload.

 We conducted an empirical study with 35 programmers of varying expertise levels performing Java code comprehension tasks. A 64-channel EEG system \cite{CompumedicsSynAmpsRT} provided direct cognitive load measurements, serving as ground truth to evaluate code region revisit behaviors captured by the Tobii Eye Tracker 5L \cite{Tobii}. 

The paper presents the following \textbf{key contributions}:

\textit{ 1) Proposes the NRevisit code understandability metric, a dynamic, human-centered metric that reflects real-time engagement and cognitive load of individual programmers in code comprehension tasks.}

\textit{ 2) Demonstrates that NRevisit correlates strongly with EEG-measured cognitive load ($r_s$ ranging from 0.9067 to 0.9860, depending on the selected NRevisit pattern), outperforming existing code understandability assessment approaches based on static code metrics.}

\textit{ 3) Validates NRevisit in predictive modeling, showing that it achieves higher $R^2$ values (outperform) compared with static complexity metrics.}

\textit{ 4) Provides practical insights for integrating NRevisit into real-world software engineering workflows, including AI-assisted code generation setups.}

The rest of this paper is structured as follows: Section 2 reviews related work on static complexity metrics, behavioral metrics and combined approaches. Section 3 details the NRevisit definitions, experimental setup and analysis methods. Section 4 presents the results on NRevisit selection and regression performance. Section 5 discusses its advantages, limitations and applications, followed by Section 6, which concludes the study and outlines future directions.
\section{Related Work}
The accurate assessment of code complexity has been a cornerstone of software engineering research, aiming to improve software quality and maintainability \cite{mccabe1976complexity}. Over the decades, various metrics have been proposed, with a few achieving widespread adoption \cite{nunez2017source}. While these metrics have proven useful in specific contexts, they often fail to capture the human-centric aspects of code comprehension, such as the cognitive load experienced by programmers \cite{kaur2019cognitive}\cite{scalabrino2017automatically}. Recent advancements in interdisciplinary research, including the integration of behavioral data and physiological measurements, have opened new avenues for addressing these limitations \cite{hao2023accuracy}\cite{peitek2022correlates}, paving the way for more dynamic and programmer specific metrics such as NRevisit, the metric proposed in this paper.

\subsection{Static Complexity Metrics} 
The study of code complexity metrics has long been recognized as vital for ensuring software quality and maintainability \cite{mccabe1976complexity}. While many metrics have been proposed \cite{nunez2017source}, only a limited number have been widely adopted in the industry. 

One of the most established metrics is \textbf{McCabe Cyclomatic Complexity (V(g))} \cite{mccabe1976complexity}, which represents the number of linearly independent paths in a program control flow.  V(g) actually represents the difficulty of testing the code (using control flow testing), but it is widely used to determine when a unit should be refactored \cite{herbold2011calculation}\cite{yamashita2016thresholds}, resulting in the simplification commonly accepted that V(g) also represents the difficulty in understanding code. In contrast, \textbf{Halstead's metrics} capture the data and vocabulary aspects of code, aiming to provide a more comprehensive view of complexity \cite{halstead1977elements}. However, these metrics are often criticized for being difficult for programmers to interpret and apply effectively \cite{curtis1984psychological}. 

More recently, \textbf{SonarSource (CC Sonar)} introduced the Cognitive Complexity metric to address the shortcomings of V(g), with the goal of offering a more intuitive measure of code understandability \cite{campbell2018cognitive}. While this metric has started gaining traction in industry settings, it still faces limitations in capturing the full range of cognitive factors involved in code comprehension \cite{hao2023accuracy}. Tools like SonarQube implement these metrics for practical applications. However, their limitations in reflecting programmer-specific cognitive demands have been widely acknowledged \cite{lavazza2023empirical}\cite{munoz2020empirical}.

\subsection{Behavioral and Neuroscience-Based Studies}
Several studies have highlighted the inherent limitations of traditional complexity metrics in addressing human-centric factors such as programmer expertise, emotional state and environmental context \cite{abbad2022estimating}\cite{aljehane2023studying}\cite{hao2023accuracy}\cite{mathur2021dynamic}. Eye-tracking studies, such as \cite{grabinger2024eye}, have demonstrated the relationship between gaze behavior and cognitive effort. Additionally, \cite{davis2022analysis} and \cite{obaidellah2018survey} conduct systematic literature reviews of eye-tracking research in software engineering, further highlighting the relevance of gaze behavior in this field. EEG-based metrics, on the other hand, provide direct measurements of cognitive load, serving as a neuroscience-based reference 
\cite{hao2023accuracy}\cite{lobo2016cognitive}\cite{medeiros2021can}.

Several studies have explored the integration of eye-tracking and EEG to understand code comprehension. For instance, \cite{ishida2019synchronized} simultaneously measured programmers' brain waves and eye movements during source code comprehension, providing insights into the cognitive processes involved. Similarly, Peitek et al. conducted a combined EEG and eye-tracking experiment to investigate programmer efficacy, highlighting the relationship between neural activity and eye movement patterns during coding tasks \cite{peitek2022correlates}.

Interdisciplinary approaches have incorporated neuroimaging and physiological measures, such as fMRI, EEG and eye-tracking, to understand the cognitive aspects of programming tasks better \cite{medeiros2021can}\cite{peitek2021program}\cite{siegmund2014understanding}. For instance, fMRI studies have examined the neural correlates of code comprehension, while EEG has been widely used to assess cognitive load during debugging and programming tasks \cite{peitek2018simultaneous}. Additionally, biometric measurements such as heart rate variability (HRV) and pupillometry have emerged as promising tools for real-time cognitive load assessment \cite{couceiro2019biofeedback}\cite{hijazi2022quality}\cite{hijazi2021ireview}\cite{muller2016using}, though their adoption in practical software development remains limited.

\section{NRevisit Method}
Since reading and comprehending code is usually not a sequential process, the intuition behind the proposed method is that the number of times a programmer revisits a code construct (region) reflects the difficulty in understanding that snippet. A high number of revisits to a given code region indicates a need for confirmation or detailed analysis, either due to the code's intrinsic complexity or the programmer's lack of experience. In either case, frequent revisits reflect a high cognitive load for that specific programmer.  

The first step in applying the proposed method is partitioning the code into basic regions. The partitioning process is straightforward and based on well-defined programming constructs such as sequences, selections and iterations. The specific boundaries of these regions were defined based on logical separations within the code, ensuring that no language constructs were broken and that each region maintains a coherent, self-contained meaning. Code region boundaries are not visible to programmers, as they serve only to measure cognitive load based on the NRevisit metric. 

We defined top-level regions as major blocks of code, such as functions or methods and sub-regions as further divisions of top-level regions based on control flow structures (e.g., loops and conditionals). Importantly, these regions do not overlap, ensuring that each region remains logically self-contained. 

\begin{figure}[h]
\vspace{-5pt} 
\centerline{\includegraphics[width=0.4\textwidth,height=0.3\textwidth]{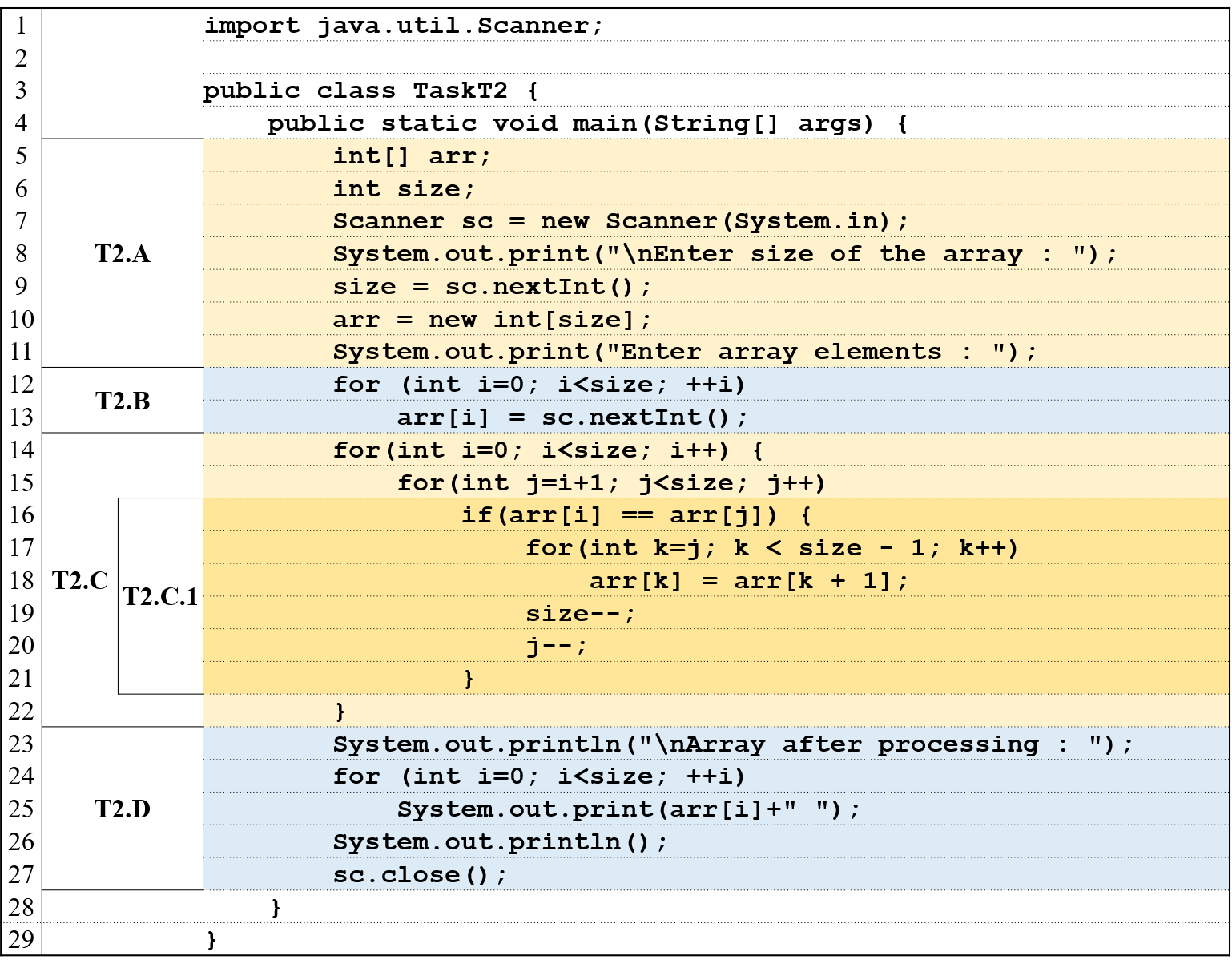}}
\caption{Code Snippet for Task 2}
\Description{Code Snippet for Task 2}
\vspace{-10pt} 
\end{figure}

To clarify the partitioning strategy, consider program T2, which is partitioned as shown in Fig.1. The code is divided into top-level regions and sub-regions, each representing a logical segment of the program. In this example:

\begin{itemize} \item \textbf{T2.A:} Defines the variable declaration and array initialization as a top-level region.
\item \textbf{T2.B:} Represents the section where the user inputs array elements.
\item \textbf{T2.C:} Includes nested loops for processing the array elements, with \item \textbf{T2.C.1} as a sub-region focusing on the inner loop logic.
\item \textbf{T2.D:} Outputs the processed array, representing the final top-level region.
\end{itemize}

Although nested regions can be used in NRevisit, we focus on top-level code regions, excluding subregions (e.g., T2.C.1), to minimize redundancy and enhance cognitive re-engagement with top-level. \textbf{Algorithm 1} details the algorithmic view of the code division steps.
\begin{algorithm}
\caption{Region identification from Source Code}
\textbf{Input:} Source code $C$ \\
\textbf{Output:} Hierarchical regions $R = \{R_1, R_2, \ldots, R_n\}$, where each $R_i$ is a top-level region or sub-region.

\begin{algorithmic}[1]
\State Parse $C$ into an Abstract Syntax Tree (AST)
\State Identify top-level regions:
    \begin{itemize}
        \item[\textbf{Rule 1:}] Every function or method definition is a top-level region $R_i$
        \item[\textbf{Rule 2:}] Global declarations (e.g., class variables, imports) form a separate top-level region
    \end{itemize}
\State Subdivide top-level regions $R_i$ into sub-regions $S_i$:
    \For{each $R_i$}
        \State \textbf{Step A:} Revisit the AST to identify control-flow boundaries (e.g., loops, conditionals)
        \State \textbf{Step B:} Partition $R_i$ into sub-regions $S_j$
        \begin{itemize}
            \item Each $S_j$ corresponds to a single control-flow element (e.g., loop, while)
            \item Nested structures create hierarchical sub-regions (e.g., $S_{j1}$ inside $S_j$)
            \item[\textbf{Rule 3:}] Sub-regions are non-overlapping, coherent, and follow the AST's hierarchical nesting
        \end{itemize}
    \EndFor
\State Assign logical boundaries:
    \begin{itemize}
        \item[\textbf{Rule 4:}] Regions are ``bounded'' or ``closed'' by syntactic delimiters (e.g., \{ \} in C/Java or indentation in Python)
        \item[\textbf{Rule 5:}] Each region should be a logical (non-overlapping), self-contained unit (e.g., function, loop, if-body)
    \end{itemize}
\State Assign labels for each region hierarchically (e.g., $R_i \rightarrow S_i \rightarrow S_{j1}$)
\end{algorithmic}
\end{algorithm}

To assess code understandability based on gaze behavior, we define \textbf{NRevisit}, a metric that quantifies how often a programmer revisits specific code regions during code comprehension. Two variants of NRevisit were evaluated, both of which rely on fixations lasting $\geq 1s$ to be considered valid:

\begin{itemize} 
     \item \textbf{Continuous Time-Based NRevisit (C NRevisit)} measures the number of distinct sustained gaze fixations within a code region. A new revisit is counted when a programmer pauses fixation on a code region and then resumes, regardless of whether their gaze shifts to another region or elsewhere on the screen before returning. 
    
    \item \textbf{Close Loop NRevisit (CL NRevisit)} counts only complete exit-return cycles, where a programmer leaves a code region to look at another region and then returns to the original region.  This filters out minor gaze shifts within the same region and focuses on full context-switching behavior. 
\end{itemize}

Figure 2 illustrates the differences between C NRevisit and CL NRevisit. The left side of the figure indicates the programmer's gaze movements while reading and understanding code, showing fixation transitions across different code regions (A, B, and C). The right table quantifies the two metrics for Region B, demonstrating how different gaze transitions contribute to each NRevisit calculation.

\begin{figure}[h]
\vspace{-10pt}
\centerline{\includegraphics[width=0.37\textwidth,height=0.17\textwidth]{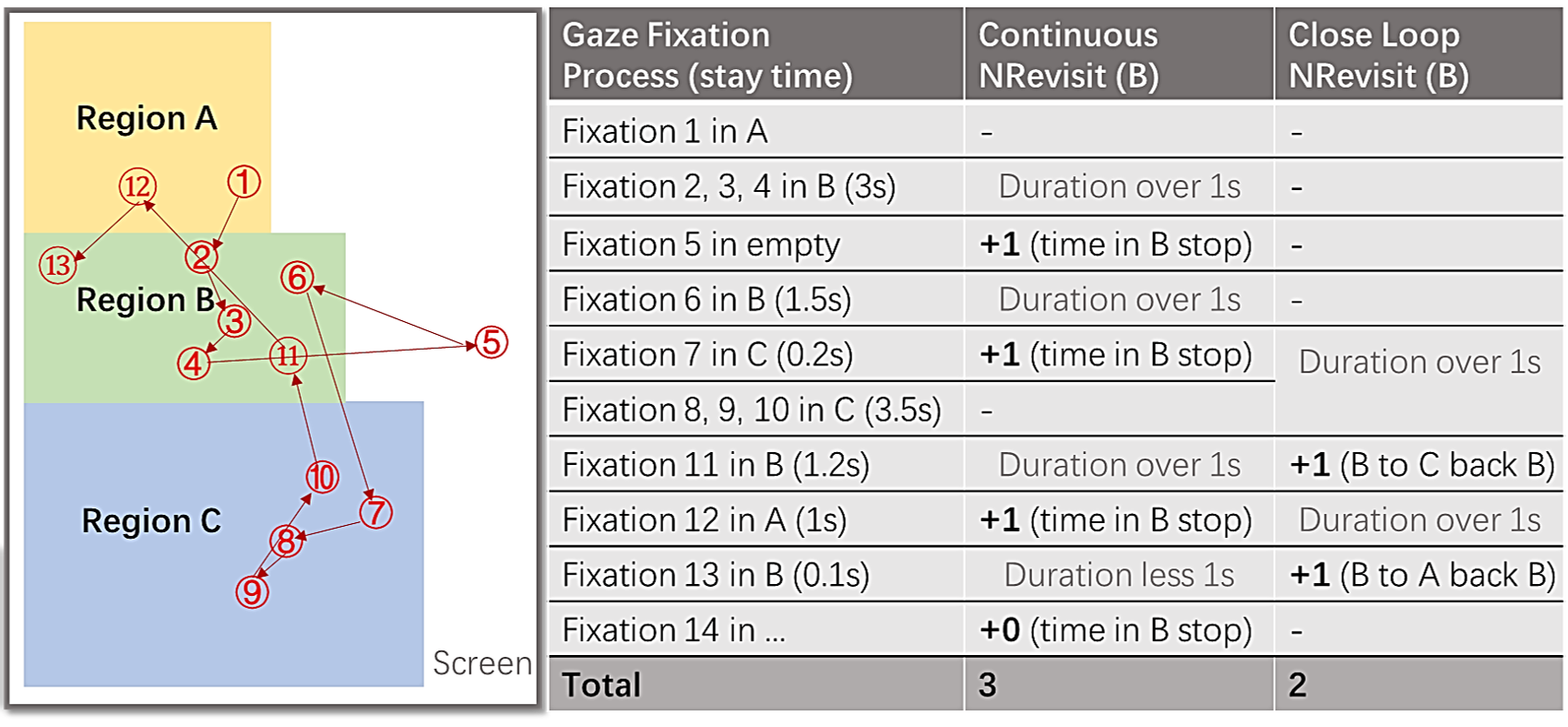}}
\caption{Example of C NRevisit and CL NRevisit in Region B}
\Description{Example of C NRevisit and CL NRevisit in Region B}
\vspace{-15pt} 
\end{figure}

\section{Experimental Evaluation}
The key idea is to measure participants' (programmers') actual cognitive load during code comprehension tasks using EEG and use these measurements as ground truth to evaluate NRevisit. 

Before the experiment, participants undergo preparation, including EEG cap attachment and eye tracker calibration. The experiment consists of four rounds, each featuring a fairy tale passage (used as baseline task) and a randomly assigned Java program from a set of four programs. Each round begins with a 30-second fixation cross task for adaptation, followed by the fary tale reading to establish a cognitive load baseline. Participants then engage in the code comprehension task for up to 10-minutes. Afterward, they completed two questionnaires to assess their code understanding, which included questions on what the code does and how the code works. The entire session lasts approximately 2 to 2.5 hours. The reproducibility package with all the data and experiment protocol can be found in the following repository: \textcolor{blue}{\href{https://github.com/SoftwareRepository2025/NReivisit}{GitHub Repository}}.

\subsection{Controlled Experiment}

\textbf{Participant Selection and Expertise Classification:}
A total of \textbf{35 participants} (28 men and 7 women) were recruited, representing diverse levels of Java programming expertise. Participants were pre-screened through interviews assessing their years of practice and lines of code (LoC) written. They were categorized as follows:
\begin{itemize} 
\item \textbf{Expert (4 participants):} Professional developers with \(\geq\) 3 years of experience.
\item\textbf{Advanced (8 participants):} Developers with \(\geq\) 1 year of experience.
\item\textbf{Intermediate (23 participants):} Students with good Java programming proficiency but without industry experience.
\end{itemize}

\textbf{Task Design and Program Complexity:}
The study involved four carefully selected Java programs (T1–T4) designed to span a range of complexity levels. All programs were implemented using common-sense algorithms and strategies, without artificially increasing complexity beyond what the task required. However, to obscure the algorithm's purpose (because the task asked to participants was to comprehend each code), variable names were neutral and the program output provided no hints about its functionality.

\textbf{Task 1}: the program prompts the user for a string and checks if it is a palindrome. The algorithm constructs a reversed version of the input string and compares both. To avoid revealing the goal, the output message deliberately omits the word \textit{palindrome}.
 
\textbf{Task 2}: the program removes duplicate values from an integer array. The algorithm scans the array, checking each element against subsequent values and removing duplicates. The array is initially populated with user input values.

\textbf{Task 3}: the code implements an iterative version of the \textbf{quicksort} algorithm to sort an array in ascending order. It uses an auxiliary array to simulate a stack for storing the lower and upper bounds of each segment being sorted. The array is first populated with user-input values and then sorted and displayed.

\textbf{Task 4}: the program simulates \textbf{Conway’s Game of Life}, where cells exist in a two-dimensional array. Each cell's survival follows predefined rules based on neighboring cells. The program generates an initial random population, computes the next generation and displays it on the screen.

To maintain practicality, the programs were limited to 50 lines of code (LoC) in total, with individual V(g) values ranging from 5 to 9 (see Table 2 in section 5). This allowed participants to complete tasks within 2 to 2.5 hours, avoiding excessive fatigue while providing sufficient variation in complexity.

The experiments were conducted in a controlled environment free of distractions or interruptions, ensuring that EEG-measured cognitive load was solely related to code comprehension.

\textbf{Apparatus and Data Collection:}
The study employed synchronized eye-tracking and EEG systems as follows: 
\begin{itemize}
\item \textbf{Eye-Tracking Device:} A Tobii Eye Tracker 5L\cite{Tobii} was used to capture fixation duration, revisit frequency and gaze transitions within predefined code regions. The device recorded gaze data at a high temporal resolution, ensuring accurate tracking of participant behavior.

\item \textbf{EEG System:} A 64-channel EEG system \cite{CompumedicsSynAmpsRT} measured cognitive load using established measured cognitive load using established features \cite{medeiros2021can}. The signals were sampled at a frequency of 1,000 Hz, and the electrodes on the scalp were positioned according to the international 10-10 system \cite{graimann2010brain}.
\end{itemize}

\textbf{Data Processing:} The signal processing pipeline for both eye-tracking and EEG data consists of three primary stages: preprocessing, feature extraction and normalization (max-min normalization was performed). Following these steps, the processed data from both modalities are aligned through a synchronization procedure to ensure temporal consistency and facilitate integrated analysis.

 \textbf{1) Eye-Tracking Data:} Table 1 shows the preprocessing for the eye-tracking signal. Gaze data was segmented based on the predefined code regions, ensuring that revisits were correctly attributed to each code region. The revisit count (NRevisit) was computed for each region using C NRevisit and CL NRevisit. Fixation durations and revisit patterns were aggregated to create participant-level behavioral summaries for statistical and regression analyses.
\begin{table}[ht]
\vspace{-10pt}
    \centering
    \renewcommand{\arraystretch}{0.9}
    \setlength{\tabcolsep}{4pt}
    \caption{Eye-Tracking Data Preprocessing Steps}
    \scalebox{0.7}{
    \begin{tabular}{p{3.8cm} p{4.8cm}}
        \hline
        \textbf{Step} & \textbf{Description} \\
        \hline
        Artifact Removal & Eliminates noise caused by blinks and sudden eye movements \cite{hansen2013pupil}. \\
        Missing Data Handling & Interpolates missing values to maintain data continuity \cite{wang2012missing}. \\
        Filtering & Applies low-pass or band-pass filters to remove high-frequency noise and smooth fluctuations \cite{boccignone2013filtering}. \\
        Coordinate Mapping & Maps gaze positions to screen coordinates for spatial consistency \cite{duchowski2007eyetracking}. \\
        Resampling & Adjusts the sampling rate to align with other physiological signals, enabling synchronization \cite{kumar2021synchronizing}. \\
        \hline
    \end{tabular}
    }
    \vspace{-10pt}
\end{table}

 \textbf{2) EEG Data:} The preprocessing of raw EEG signals followed established standards in EEG research \cite{bigdely2015prep}, incorporating artifact removal techniques to enhance signal quality. Specifically, bandpass filtering was applied to eliminate unwanted frequency components, while independent component analysis (ICA) was utilized to reduce noise artifacts caused by blinks and muscle activity.

After preprocessing the raw EEG data, key features were extracted to quantify cognitive load (CL). Specifically, \( EEG1 \) from electrode \( \text{F2} \) and \( EEG2 \) from electrode \( \text{PZ} \) were selected based on established methodologies \cite{medeiros2021can}.

\vspace{-1pt}
\begin{equation}
    EEG1 = \frac{\theta}{\beta + \alpha}
\end{equation}
\vspace{-1pt}
\begin{equation}
    EEG2 = \frac{\theta}{\alpha}
\end{equation}

where \textbf{ \( \theta \)} (4–8 Hz) is linked to memory processing and cognitive workload, \textbf{\( \alpha \)} (8–12 Hz) to relaxed wakefulness and \textbf{ \( \beta \)} (13–30 Hz) to active problem-solving.

These two features were quantified using the area under the curve during eye-tracked fixations on code regions, as the final cognitive load. Repeated visits were summed to assess the cumulative cognitive load, following established EEG-based cognitive load analysis methods \cite{medeiros2021can}. For example, a cognitive load of 20 while understanding code indicates that the programmer's cognitive load was 20 times higher than the load measured in the control task (fairy tale reading in the native language of participants). 

 \textbf{3) Synchronization of Eye-Tracking and EEG Data:} Time stamps from the eye tracker and EEG system were matched to ensure that revisit events and cognitive load measurements corresponded to the same moments during task execution. This synchronization enabled the identification of high-cognitive-load regions based on both behavioral and physiological data.

\subsection{Analysis Approach}
This study employs a multi-step analysis approach to assess the effectiveness of NRevisit in predicting the actual cognitive load measured by EEG (CL EEG). The evaluation consists of correlation analysis, feature selection and regression modeling.

\textbf{Correlation Analysis Between NRevisit and Cognitive Load:} Understanding the relationship between NRevisit values and EEG-measured cognitive load requires an assessment of both monotonic and linear dependencies.
\begin{itemize}
\item \textbf{Spearman’s correlation \cite{spearman1904proof}} coefficient ($r_s$, $\rho$) measures rank-based relationships, ensuring robustness against nonlinearity in the distribution of cognitive load.
\item \textbf{Pearson’s correlation \cite{pearson1920history}} coefficient (R) and determination coefficient ($R^2$) evaluate the linear dependence of NRevisit on CL EEG, determining its direct predictive power.
\end{itemize}

\textbf{Regression Modeling for Cognitive Load Prediction:}  
A comparative analysis of predictive models demonstrates the effectiveness of CL NRevisit over static complexity metrics (V(g), HEff, HDif, CC Sonar). To evaluate the model’s generalizability and mitigate potential biases, we adopted a random selection approach, allocating approximately 30\% of subjects to the testing set and 70\% to the training set. The following models were employed:  
\begin{itemize}
    \item \textbf{Linear Regression (LR)} provides a baseline for how well each metric predicts CL EEG in a simple linear setting.
    \item \textbf{Neural Networks (NN)} capture non-linear relationships between revisit patterns and cognitive load. A feedforward NN with a single hidden layer, trained using a gradient-based optimization algorithm to capture non-linear relationships.
    \item \textbf{Gaussian Process Regression (GPR)} model with an automatic relevance determination (ARD) kernel models probabilistic dependencies, providing robust uncertainty estimates and ensuring a reliable evaluation of prediction uncertainty.
\end{itemize}

Each model was evaluated over 100 randomizations to ensure robust performance metrics. We recorded the \textbf{maximum \(R^2\)}, \textbf{minimum MSE} and \textbf{average \(R^2\)} to assess the predictive strength of each model. Additionally, the \textbf{standard deviation of \(R^2\) and MSE} was calculated to measure the stability and consistency of the models across different data splits. The \(R^2\) (coefficient of determination) and MSE (mean squared error) are defined as:
\begin{equation}
  \ MSE = \frac{1}{n}\sum_{i=1}^{n}(y - \hat{y})^{2}
\end{equation}
\vspace{-7pt}
\begin{equation}
  \ R^{2} = 1 - \frac{{\sum (y - \hat{y})}^{2}}{{\sum (y - \bar{y})}^{2}}
\end{equation}
where \(y_i\) is the true value, \(\hat{y}_i\) is the predicted value, \(\bar{y}\) is the mean of the true values and \(n\) is the number of samples.

\textbf{Feature Selection and Predictive Importance Analysis:} Identifying the most informative predictors of cognitive load involves a structured feature selection process using machine learning models. Candidate features include V(g), HEff, HDif, CC Sonar (static complexity metrics) and NRevisit (behavioral metric). Each model runs across five selection rounds with 20 randomizations per round, ensuring stability in identifying the most influential features.

To address feature selection challenges in our small datasets, we employed random data splitting, cross-validation and sequential feature selection (SFS) \cite{whitney1971direct}. Random splitting mitigates overfitting by evaluating models on multiple randomized test sets, while 5-fold cross-validation reduces bias and enhances evaluation stability. SFS, combined with regression models, iteratively identified the most relevant features based on Mean Squared Error (MSE) and R² metrics. The number of times each feature is selected over 100 runs, along with its frequency in high-performing models ($R^2 \geq 0.8$), determines its contribution to cognitive load prediction.

\section{Results and Analysis}
Table 2 presents the scores for the four whole programs on static complexity metrics, EEG-measured cognitive load, and dynamic gaze-based revisit metrics, the latter two being the sum of top-level code regions for each program. Given the controlled conditions of the experiment, free of external stimuli or interruptions, the cognitive load measured by EEG reflects the mental effort required to comprehend each program. It is evident that code complexity metrics do not fully align with the average EEG-measured cognitive load (across all participants). For example, \textbf{T2} has a much higher CC Sonar value than \textbf{T1}, but the EEG cognitive load does not reflect this difference, especially for expert and advanced. Similarly, \textbf{T4} is the most complex program according to all static metrics, yet \textbf{T3} recorded the highest average cognitive load which is 50.42. 

Further analysis of programmers' behavior across expertise levels aligns with the strong positive correlation between EEG load and NRevisit. Intermediate programmers exhibit more frequent and effortful revisits, leading to a higher EEG cognitive load. For instance, in task T2, intermediates show a C NRevisit of 71 and CL NRevisit of 35, with corresponding EEG loads of 30.35 and 24.86—both higher than those of experts (C NRevisit: 47, CL NRevisit: 22). In contrast, experts adopt a more structured reading strategy, minimizing unnecessary revisits and maintaining lower EEG load. This pattern is also evident in task T4, where experts have C NRevisit and CL NRevisit values of 32 and 27, respectively, compared to 112 and 45 for intermediates. These findings further confirm that intermediates face greater cognitive challenges when processing complex code, while experts navigate more efficiently.
\begin{table*}[]
\centering
\setlength{\tabcolsep}{4pt} %
\renewcommand{\arraystretch}{1.2} %
\caption{Code Complexity Measured by Static and Dynamic Metrics, as well as CL EEG}
\scalebox{0.7}{ 
\begin{tabular}{
    c c c c c c 
    >{\columncolor[HTML]{DDEBF7}}c
    >{\columncolor[HTML]{DDEBF7}}c
    >{\columncolor[HTML]{DDEBF7}}c
    >{\columncolor[HTML]{DDEBF7}}c
    >{\columncolor[HTML]{FFF2CC}}c
    >{\columncolor[HTML]{FFF2CC}}c
    >{\columncolor[HTML]{FFF2CC}}c
    >{\columncolor[HTML]{FFF2CC}}c
    >{\columncolor[HTML]{E2EFDA}}c
    >{\columncolor[HTML]{E2EFDA}}c
    >{\columncolor[HTML]{E2EFDA}}c
    >{\columncolor[HTML]{E2EFDA}}c
}
\hline
& & & & & &
\multicolumn{4}{c}{\cellcolor[HTML]{BDD7EE}\textbf{Cognitive load (EEG)}} &
\multicolumn{4}{c}{\cellcolor[HTML]{FFD966}\textbf{C NRevisit}} &
\multicolumn{4}{c}{\cellcolor[HTML]{A9D08E}\textbf{CL NRevisit}} \\ 
\cline{7-18}
\multirow{-2}{*}{\textbf{Region}} & 
\multirow{-2}{*}{\textbf{LoC}} & 
\multirow{-2}{*}{\textbf{V(g)}} & 
\multirow{-2}{*}{\textbf{HEff}} & 
\multirow{-2}{*}{\textbf{HDif}} & 
\multirow{-2}{*}{\textbf{\begin{tabular}[c]{@{}c@{}}CC\\ Sonar\end{tabular}}} &
\textbf{All} & \textbf{Exp.} & \textbf{Adv.} & \textbf{Inter.} & 
\textbf{All} & \textbf{Exp.} & \textbf{Adv.} & \textbf{Inter.} & 
\textbf{All} & \textbf{Exp.} & \textbf{Adv.} & \textbf{Inter.} \\ 
\hline
\textbf{T1} & 26 & 5 & 20035.69 & 23.69 & 6 &
22.21 & 21.04 & 23.73 & 25.16 &
62 & 47 & 52 & 71 &
30 & 22 & 29 & 35 \\ 
\hline
\textbf{T2} & 27 & 7 & 29140.08 & 28 & 12 &
28.53 & 24.86 & 25.18 & 30.35 &
63 & 50 & 60 & 72 &
34 & 31 & 32 & 42 \\ 
\hline
\textbf{T3} & 49 & 8 & 84543.6 & 45.72 & 12 &
50.42 & 41.68 & 51.14 & 65.76 &
134 & 117 & 122 & 138 &
46 & 37 & 43 & 52 \\ 
\hline
\textbf{T4} & 35 & 9 & 128663.8 & 63.63 & 17 &
44.48 & 29.87 & 42.58 & 47.44 &
101 & 32 & 73 & 112 &
44 & 27 & 43 & 45 \\ 
\hline
\end{tabular}
}
\vspace{-9pt}
\end{table*}

Figure 3 and Figure 4 illustrate the relationship between gaze-based revisit metrics (CL NRevisit and C NRevisit) and EEG-measured cognitive load (CL EEG) across the code regions of the top level of the four programs (T1 to T4). The values represent the average of all participants. The blue solid line represents the NRevisit metric, while the red dashed line denotes CL EEG, providing insights into the cognitive effort associated with different segments of the code. 

Both CL NRevisit and C NRevisit exhibit a high positive correlation with CL EEG, showing that higher revisit rates are associated with increased cognitive processing effort. Peaks in both NRevisit metrics align with increased CL EEG values, suggesting regions with frequent gaze revisits correspond to higher cognitive demands.

\vspace{-5pt}
\begin{figure}[h]
\centerline{\includegraphics[scale=0.32]{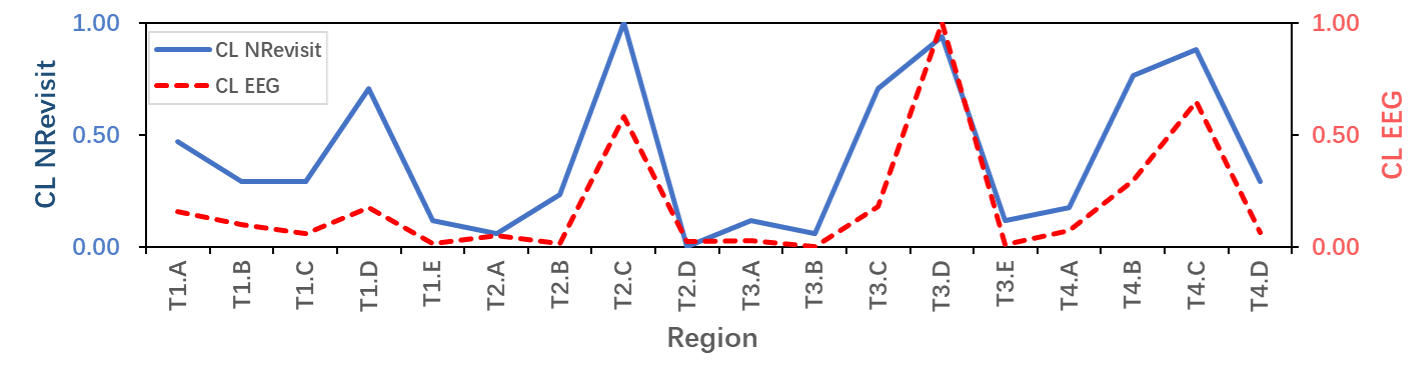}}
\vspace{-5pt} 
\caption{CL EEG versus CL NRevisit}
\Description{CL EEG versus CL NRevisit}
\end{figure}

\begin{figure}[h]
\vspace{-25pt} 
\centerline{\includegraphics[scale=0.32]{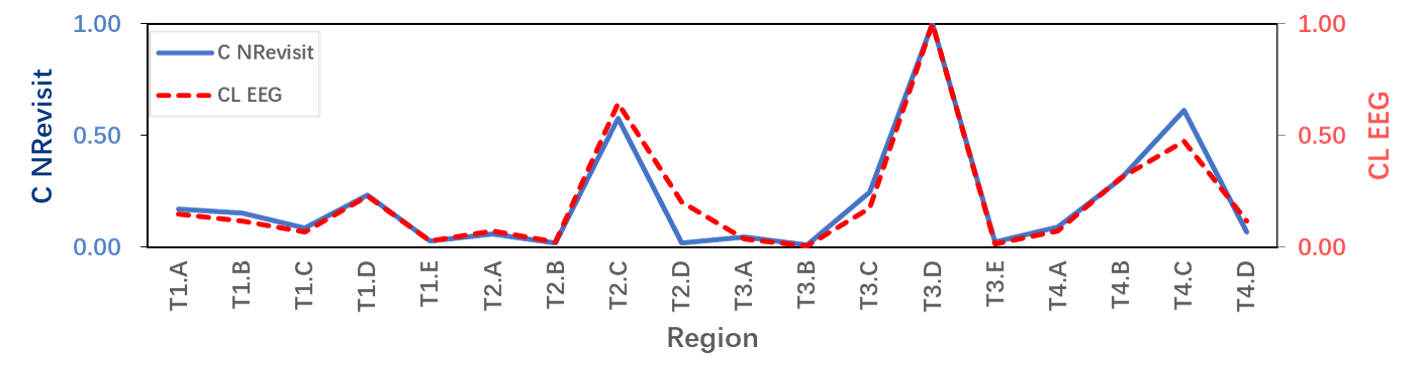}}
\vspace{-5pt} 
\caption{CL EEG versus C NRevisit}
\Description{CL EEG versus C NRevisit}
\vspace{-15pt} 
\end{figure}

To further examine the relationship between the revisit metric and EEG-measured cognitive load, we conducted a correlation analysis (Table 3). This analysis uses monotonic and linear dependencies, considering different expertise levels to assess how programming proficiency influences revisit behavior and cognitive processing demands. Two significant insights emerged from this analysis:

\begin{table}[]
    \centering
    \small
    \renewcommand{\arraystretch}{0.9}
    \newcommand{\hlgreen}{\rowcolor[rgb]{0.75,1,0.75}}

    \caption{Spearman and Pearson Correlation Results}
    
    \scalebox{0.75}{
    \begin{tabular}{lcc|cc}  
        \toprule
        \textbf{Expert level (Num.)} & \multicolumn{2}{c}{\textbf{Spearman correlation}} & \multicolumn{2}{c}{\textbf{Pearson correlation}} \\
        \midrule
        \hlgreen C NRevisit (defined by continuous time) & $r_s$ & $\rho$ (2-tailed) & R & $R^2$ \\
        \midrule
        Expert (4)       & 0.8640  & $\approx 0$  & 0.9262  & 0.8578 \\
        Advanced (8)     & 0.9565  & $\approx 0$  & 0.9243  & 0.8543 \\
        Intermediate (23) & 0.9876  & $\approx 0$  & 0.9852  & 0.9706 \\
        Global (35)      & 0.9860  & $\approx 0$  & 0.9966  & 0.9932 \\
        \midrule
        \hlgreen CL NRevisit (defined by close loop) & $r_s$ & $\rho$ (2-tailed) & R & $R^2$ \\
        \midrule
        Expert (4)       & 0.8395  & $0.00017$   & 0.6680  & 0.4462 \\
        Advanced (8)     & 0.8483  & $1 \times 10^{-5}$  & 0.7189  & 0.5168 \\
        Intermediate (23) & 0.8708  & $\approx 0$  & 0.8597  & 0.7391 \\
        Global (35)      & 0.9067  & $\approx 0$  & 0.8435  & 0.7115 \\
        \bottomrule
    \end{tabular}
    }
    \vspace{-9pt}
\end{table}

\textbf{1) NRevisit: A Comprehensive and Reliable Measure of Cognitive Load.} For the global level, both C NRevisit and CL NRevisit show strong correlation values with the cognitive load measured by EEG. C NRevisit consistently achieves higher correlation values compared to CL NRevisit ($r_s$ = 0.9067, $R^2$ = 0.7115), which reaches near-perfect correlation values ($r_s$ = 0.9860, $R^2$ = 0.9932), indicating a stronger association with cognitive load. This could be due to the closed-loop nature of CL NRevisit, which may not capture as many fine-grained variations as the C NRevisit. 

The Spearman correlation indicates a strong monotonic relationship between NRevisit and EEG-measured cognitive load. The high Pearson correlation suggests a very strong linear relationship between NRevisit and cognitive load.

\textbf{2) Expertise-Level Differences in NRevisit Metrics.} For the data of \textbf{experts} show lower correlations in both NRevisit ($r_s$ = 0.8640 and 0.8395, $R^2$ = 0.8578 and 0.4462), reflecting that it isn't always the case that higher cognitive efficiency leads to higher revisit rates, as experts may have fewer unnecessary re-engagements. Compared with other levels (except all), \textbf{intermediate} participants display the highest correlation values ($r_s$ = 0.9876 and 0.8708, $R^2$ = 0.9706 and 0.7391), suggesting that NRevisit aligns well with their iterative reading patterns. 

These findings highlight that NRevisit effectively differentiates expertise levels and aligns better with cognitive processing differences, as it captures active comprehension struggles rather than passive viewing.

\subsection{Feature Selection and Regression Performance}
Table 4 presents the regression performance of NRevisit, as well as the performance of the four static complexity metrics. The CL EEG was used as ground truth. The comparison spans Linear Regression (LR), Neural Networks (NN) and Gaussian Process Regression (GPR), each tested over 100 randomizations to evaluate robustness. We obtained the best R² and MSE values from the 100 randomizations, along with their average values and the corresponding standard deviation. This provides insights into both the optimal performance and consistency of each model.

\begin{table}[h]
\vspace{-5pt}
\caption{Regression Performance Across 100 Randomizations}
\scalebox{0.65}{
\begin{tabular}{cccccccc}
\hline
\multicolumn{8}{c}{\textbf{100 randomizations}}                                                            \\ \hline
 &                                    & \textbf{Max} & \textbf{Min} & \multicolumn{4}{c}{\textbf{Average}} \\ \cline{3-8} 
\multirow{-2}{*}{\textbf{Regression Model}} &
  \multirow{-2}{*}{\textbf{Metric}} &
  $R^2$ &
  MSE &
  $R^2$ &
  MSE &
  \multicolumn{2}{c}{\textbf{std ($R^2$/MSE)}} \\ \hline
 & V(g)                               & 0.6798       & 0.0333       & 0.5753  & 0.0425  & 0.0581  & 0.0058 \\ \cline{2-8} 
 & HEff                                & 0.7602       & 0.0239       & 0.6566  & 0.0334  & 0.0620  & 0.0054\\ \cline{2-8} 
 & HDif                               & 0.7140       & 0.0282       & 0.6265  & 0.0372  & 0.0447  & 0.0050  \\ \cline{2-8} 
 & CC sonar                           & 0.7118       & 0.0300       & 0.6228  & 0.0376  & 0.0480  & 0.0044 \\ \cline{2-8} 
 & C NRevisit & \textbf{0.8745}      & \textbf{0.0116}  & \textbf{0.7722}  & \textbf{0.0226}  & \textbf{0.0488}  & \textbf{0.0048} \\ \cline{2-8} 
\multirow{-6}{*}{\textbf{Linear regression}} &
  CL NRevisit & 0.6540 & 0.0356 & 0.5841 & 0.0417 & 0.0351 & 0.0032 \\ \hline
 & V(g)                               & 0.7271       & 0.0240       & 0.6922  & 0.0304  & 0.0571  & 0.0053 \\ \cline{2-8} 
 & HEff                               & 0.7939       & 0.0215       & 0.7323  & 0.0266  & 0.0474  & 0.0048 \\ \cline{2-8} 
 & HDif                               & 0.8084      & 0.0198       & 0.7189  & 0.0280  & 0.0552  & 0.0053 \\ \cline{2-8} 
 & CC sonar                           & 0.7670  & 0.0235       & 0.6995  & 0.0300  & 0.0409  & 0.0038 \\ \cline{2-8} 
 & C NRevisit & \textbf{0.8590}       & \textbf{0.0130}  & \textbf{0.7714}  & \textbf{0.0227}  & \textbf{0.0495}  & \textbf{0.0047} \\ \cline{2-8} 
\multirow{-6}{*}{\textbf{\begin{tabular}[c]{@{}c@{}}NN   regression\\      (Neural Networks)\end{tabular}}} & CL NRevisit & 0.8513 & 0.0164 & 0.7434 & 0.0252 & 0.0461 & 0.0046 \\ \hline
 & V(g)                               & 0.8107       & 0.0204       & 0.7313  & 0.0267  & 0.0480  & 0.0049 \\ \cline{2-8} 
 & HEff                               & 0.8374       & 0.0172       & 0.7485  & 0.0252  & 0.0557  & 0.0055 \\ \cline{2-8} 
 & HDif                               & 0.8191       & 0.0183        & 0.7388  & 0.0260  & 0.0503  & 0.0049 \\ \cline{2-8} 
 & CC sonar                           & 0.8127       & 0.0188       & 0.7280  & 0.0265  & 0.0533  & 0.0049 \\ \cline{2-8} 
 & C NRevisit & \textbf{0.8595}       & \textbf{0.0130}   & \textbf{0.7713}  & \textbf{0.0227}  & \textbf{0.0491}  & \textbf{0.0047} \\ \cline{2-8} 
\multirow{-6}{*}{\textbf{\begin{tabular}[c]{@{}c@{}}Gaussian Process   \\ Regression\end{tabular}}} & CL NRevisit & 0.8491 & 0.0168 & 0.7527 & 0.0244 & 0.0480 & 0.0044 \\ \hline
\end{tabular}}
\vspace{-10pt}
\end{table}

We discuss it from two key points:

\textbf{1) Performance of NRevisit Across Models.} The performance of NRevisit varies significantly across different models. Compared with CL NRevisit, C NRevisit consistently outperforms across all models. In \textbf{LR model)}, CL NRevisit has the lowest average $R^2$ = 0.6540, suggesting that a simple linear model fails to capture its relationship with CL EEG adequately. In contrast, C NRevisit achieves the highest performance, with a maximum $R^2$ of 0.8745 and an average of 0.7722, indicating a stronger linear association. This discrepancy likely stems from the nature of their definitions: C NRevisit, being continuously measured, retains richer temporal information, making it inherently more suited for linear modeling compared to the discrete, event-based structure of CL NRevisit.
    
Both \textbf{NN} and \textbf{GPR} models demonstrate strong performance in capturing the non-linear relationship between Nrevisit and CL EEG, highlighting their effectiveness in modeling complex cognitive behaviors. CL NRevisit achieves a maximum $ R^2$ of 0.8513 with an average of 0.7434 in NN, and 0.8491 with an average of 0.7527 in GPR, highlighting the models' effectiveness in modeling complex relationships. Additionally, C NRevisit performs slightly better, reaching a maximum $R^2$ of 0.8591 and an average of 0.7714 in NN, and a maximum $R^2$ of 0.8595 with an average of 0.7713 in GPR. 
    
\textbf{2) Comparison with Traditional Complexity Metrics.} Non-linear models (NN and GPR) outperform traditional complexity metrics across all metrics. Among the traditional metrics, Halstead Metrics (HEff, HDif) outperform V(g) and CC Sonar, with HEff achieving a maximum $R^2$ of 0.8374 and an average of 0.7485, compared to CC Sonar's maximum of 0.8127 and an average of 0.7280. This could indicate that Halstead Metrics, by incorporating more comprehensive code features, are better at reflecting cognitive load than V(g) and CC Sonar. However, it is important to note that NRevisit outperforms all traditional metrics, achieving a maximum $R^2$ of 0.8591 and an average of 0.7714 in NN, and a maximum $R^2$ of 0.8595 and an average of 0.7713 in GPR. 

Both C NRevisit and CL NRevisit metrics significantly outperformed classical metrics (V(g), HEff, HDif, and CC sonar) in terms of explaining the variance in cognitive load (difficulty of comprehension) as measured by R² (p = 0.0025 to 0.0028), as well as in terms of prediction accuracy, measured by MSE (p = 0.0024 to 0.0028).

\begin{quote}
\vspace{-3pt}
\colorbox{yellow!10}{\begin{minipage}[t]{0.41\textwidth} 
\textbf{\textit{Conclusion 1:}} \textit{NRevisit is a strong cognitive behavioral metric, as its predictive power is maximized in non-linear models (NN and GPR); Traditional complexity metrics do not sufficiently reflect cognitive load, making NRevisit a more reliable alternative.}
\end{minipage}}
\vspace{-5pt}
\end{quote}

\subsection{NRevisit in GPR and NN Models}
We further conducted feature selection experiments using Neural Networks (NN) and Gaussian Process Regression (GPR), performing five selection rounds with 20 randomizations each round. Table 5 and Table 6 show the best results obtained regarding $R^2$ and MSE when selecting CL NRevisit and C NRevisit, assessing the frequency with which NRevisit is selected as a significant feature and whether it contributes to high-performing models ($R^2 \geq 0.8$).

\textbf{NRevisit is the Dominant Feature in NN and GPR Models.}
Across five rounds of selection with 20 randomizations per round, both NRevisit metrics are consistently chosen as the primary feature when achieving the highest $R^2$ and lowest MSE in both NN and GPR models. Features such as V(g), HEff, HDif and CC Sonar were never selected in cases. This suggests that NRevisit provides the most valuable information for predicting cognitive load compared to traditional static complexity metrics.

\begin{table}[h]
\vspace{-8pt}
\renewcommand{\arraystretch}{0.7}
\caption{Feature Selection Result with CL NRevisit}
\scalebox{0.7}{
\begin{tabular}{ccccc}
\hline
\multicolumn{5}{c}{\cellcolor[HTML]{E7E6E6}\begin{tabular}[c]{@{}c@{}}Each round is 20 randomizations\\      Features: {[}1{]}V(g), {[}2{]}HEff,   {[}3{]}HDif, {[}4{]}CC Sonar, {[}5{]}\textbf{CL NRevisit}\end{tabular}} \\ \hline
\multicolumn{5}{c}{\cellcolor[HTML]{92D050}\textbf{NN}}  \\ \hline
\textbf{\begin{tabular}[c]{@{}c@{}}Select \\  Rounds\end{tabular}} &
  \textbf{\begin{tabular}[c]{@{}c@{}}Feature\\ selected\end{tabular}} &
  \textbf{MSE} &
  $R^2$ &
  \textbf{\begin{tabular}[c]{@{}c@{}}Cumulative feature\\ importance\end{tabular}} \\ \hline
1      & {[}5{]}     & 0.0172    & 0.8329    & 0.0004    \\ 
2      & {[}5{]}     & 0.0172    & 0.8440    & 0.0004    \\ 
3      & {[}5{]}     & 0.0154    & 0.8535    & 0.0004    \\ 
4      & {[}5{]}     & 0.0192    & 0.8171    & 0.0004    \\ 
5      & {[}5{]}     & 0.0177    & 0.8402    & 0.0004    \\ \hline
\multicolumn{5}{l}{\begin{tabular}[c]{@{}l@{}}\textbf{Note:} Cumulative Feature Importance represents the total contribution of \\ selected features to the model’s   predictions.\end{tabular}} \\ \hline
\multicolumn{5}{c}{\cellcolor[HTML]{92D050}\textbf{GRP}} \\ \hline
\textbf{\begin{tabular}[c]{@{}c@{}}Select\\ Rounds\end{tabular}} &
  \textbf{\begin{tabular}[c]{@{}c@{}}Feature\\ selected\end{tabular}} &
  \textbf{MSE} &
  $R^2$ &
  \textbf{\begin{tabular}[c]{@{}c@{}}Length Scale\end{tabular}} \\ \hline
1      & {[}5{]}     & 0.0181    & 0.8262    & 0.4062    \\
2      & {[}5{]}     & 0.0158    & 0.8349    & 0.0637    \\ 
3      & {[}5{]}     & 0.0185    & 0.8285    & 0.0653    \\ 
4      & {[}5{]}     & 0.0174    & 0.8387    & 0.0630    \\ 
5      & {[}5{]}     & 0.0211    & 0.8147    & 0.4182    \\ \hline
\multicolumn{5}{l}{\begin{tabular}[c]{@{}l@{}}\textbf{Note:} Length Scale determines how smooth the predictions are in\\ the Gaussian Process Regression model.\end{tabular}} \\ \hline
\end{tabular}}
\vspace{-15pt}
\end{table}

%
\begin{table}[h]
\renewcommand{\arraystretch}{0.7}
\caption{Feature Selection Result with C NRevisit}

\scalebox{0.7}{
\begin{tabular}{ccccc}
\hline
\multicolumn{5}{c}{\cellcolor[HTML]{E7E6E6}\begin{tabular}[c]{@{}c@{}}Each round is 20 randomizations\\      Features: {[}1{]}V(g), {[}2{]}HEff,   {[}3{]}HDif, {[}4{]}CC Sonar, {[}6{]}\textbf{C NRevisit}\end{tabular}} \\ \hline
\multicolumn{5}{c}{\cellcolor[HTML]{92D050}\textbf{NN}}  \\ \hline
\textbf{\begin{tabular}[c]{@{}c@{}}Select \\  Rounds\end{tabular}} &
  \textbf{\begin{tabular}[c]{@{}c@{}}Feature\\ selected\end{tabular}} &
  \textbf{MSE} &
  $R^2$ &
  \textbf{\begin{tabular}[c]{@{}c@{}}Cumulative feature\\ importance\end{tabular}} \\ \hline
1      & {[}6{]}     & 0.0159    & 0.8372    & 0.0003    \\ 
2      & {[}6{]}     & 0.0160    & 0.8481    & 0.0003    \\ 
3      & {[}6{]}     & 0.0151    & 0.8373    & 0.0003    \\ 
4      & {[}6{]}     & 0.0162    & 0.8369    & 0.0003    \\ 
5      & {[}6{]}     & 0.0160    & 0.8479    & 0.0003    \\ \hline
\multicolumn{5}{c}{\cellcolor[HTML]{92D050}\textbf{GRP}} \\ \hline
\textbf{\begin{tabular}[c]{@{}c@{}}Select\\ Rounds\end{tabular}} &
  \textbf{\begin{tabular}[c]{@{}c@{}}Feature\\ selected\end{tabular}} &
  \textbf{MSE} &
  $R^2$ &
  \textbf{\begin{tabular}[c]{@{}c@{}}Length Scale\end{tabular}} \\ \hline
1      & {[}6{]}     & 0.0156    & 0.8449    & 0.9170    \\
2      & {[}6{]}     & 0.0174    & 0.8332    & 0.8302    \\ 
3      & {[}6{]}     & 0.0158    & 0.8541    & 0.9533    \\ 
4      & {[}6{]}     & 0.0153    & 0.8467    & 0.7466    \\ 
5      & {[}6{]}     & 0.0146    & 0.8508    & 0.8388    \\ \hline
\end{tabular}}
\vspace{-20pt}
\end{table}

In the \textbf{GPR model}, the average \( R^2 \) ranges from 0.8147 to 0.8387 for CL NRevisit, C NRevisit is 0.8332 to 0.8541, demonstrating strong predictive capability, while MSE remains low (CL NRevisit: 0.0158–0.0211, C NRevisit: 0.0146 -0.0174), confirming their contribution to stable, low-error predictions. The length scale (C NRevisit: 0.8302-0.9533, CL NRevisit: 0.0637–0.4182) further indicates its sensitivity to fine-grained variations in cognitive complexity, indicating that NRevisit captures subtle changes in cognitive load. 

In the \textbf{NN model}, despite NRevisit's relatively low cumulative feature importance (0.0004 for CL NRevisit and 0.0003 for C NRevisit), it is still prioritized for selection and strong predictive capability (CL NRevisit: 0.8171–0.8535, C NRevisit: 0.8369-0.8481). This indicates that NRevisit contains valuable information for predicting cognitive load, even if its direct importance is not the highest. 

Figures 5 and 6 illustrate the selection frequency of individual features across 100 runs. In \textbf{Figure 5}, HEff is the most frequently selected feature in both GPR (68 times) and NN (70 times). CL NRevisit is also frequently selected, though less often than HEff (32 times in GPR, 30 times in NN). V(g) and CC Sonar were not selected, suggesting that traditional control-flow-based complexity measures do not strongly correlate with cognitive load. Although CL NRevisit was selected less frequently than HEff, the number of cases with $R^2 \geq 0.8$ was higher for CL NRevisit. In GPR, CL NRevisit achieved 
$R^2 \geq 0.8$ in 11 cases, compared to only 5 cases for HEff. Similarly, in NN, CL NRevisit reached this threshold in 10 cases, whereas HEff did so in only 4. This suggests that, despite being chosen less often, CL NRevisit plays a crucial role in enhancing predictive accuracy.

\begin{figure}[h]
\vspace{-5pt}
\centerline{\includegraphics[width=0.35\textwidth,height=0.1\textwidth]{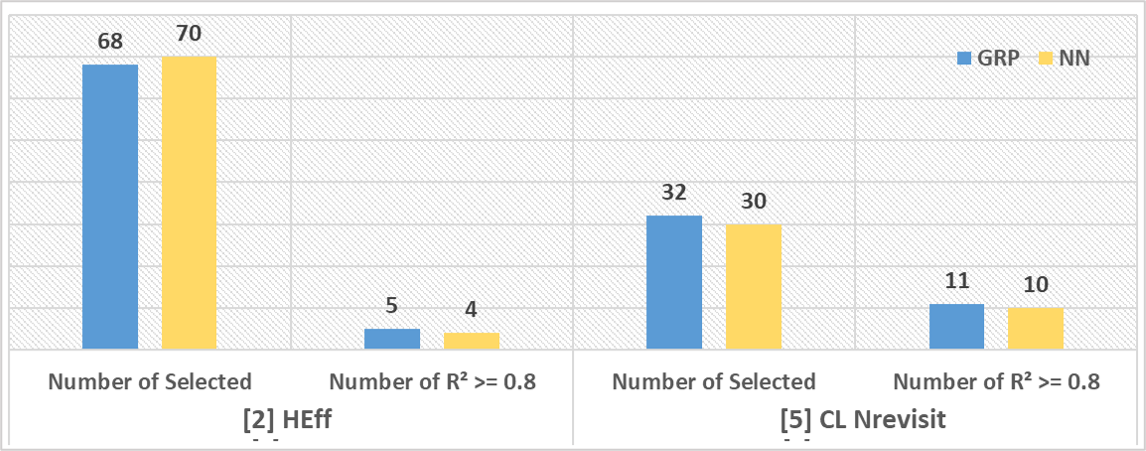}}
\caption{Selection Frequency with CL NRevisit}
\Description{Selection Frequency with CL NRevisit}
\end{figure}

\begin{figure}[h]
\vspace{-10pt}
\centerline{\includegraphics[width=0.35\textwidth,height=0.1\textwidth]{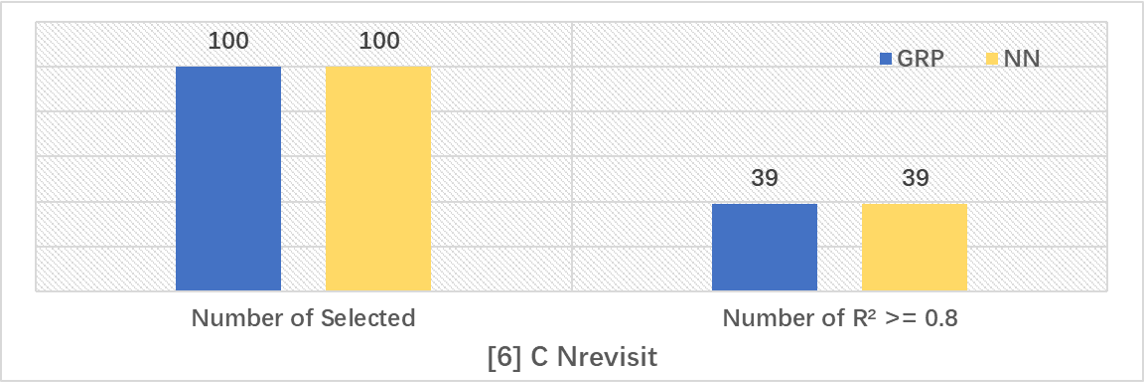}}
\caption{Selection Frequency with C NRevisit}
\Description{Selection Frequency with C NRevisit}
\vspace{-10pt}
\end{figure}

C NRevisit was selected in all 100 runs across both models, as shown in \textbf{Figure 6}, highlighting its significance in the model selection process. However, among these runs, only 39 cases achieved $R^2 \geq 0.8$. This suggests that while C NRevisit is consistently recognized as an important feature, its selection alone does not guarantee high predictive accuracy. 

\begin{quote}
\colorbox{yellow!10}{\begin{minipage}[t]{0.41\textwidth} 
\textbf{\textit{Conclusion 2:}} \textit{ \textbf{NRevisit Provides Unique Cognitive Complexity Insights.} Traditional complexity metrics such as HEff, V(g) and CC Sonar focus on static code properties. In contrast, NRevisit captures real-time cognitive engagement, providing behavioral context that static structural metrics lack.}
\end{minipage}}
\end{quote}

\section{Discussion}
Due to its simplicity, \textbf{NRevisit} has strong applicability beyond research settings, but it also has challenges and limitations.. The following subsections analyze these aspects in detail.

\subsection{Advantages of NRevisit}
The advantages of NRevisit over static complexity metrics are:

\textbf{Personalized Code Understandability Assessment:} NRevisit adapts to the cognitive processing of individual programmers, offering a dynamic complexity score that reflects real-time comprehension difficulties of individual programmers. 

\textbf{Real-Time Complexity Analysis with Low-Cost Hardware:} NRevisit can be computed using low-cost desktop eye trackers (USD 100-200) or webcam-based eye-tracking, making it affordable and scalable for real-world use. The resolution required for counting gaze revisits is significantly lower than full eye-tracking analytics, allowing for practical integration into IDEs and developer tools.

\textbf{Identification of Cognitively Demanding Code Regions:} NRevisit directly highlights code areas that are frequently revisited, providing a more practical and actionable understanding of which parts of the code are cognitively demanding and may require refactoring, improved development process.

\textbf{Capturing Complexity as Experienced by Developers:}
 While NRevisit can only be measured after programmers' interaction with the code, it reflects real-world cognitive challenges more accurately than static metrics. As such, NRevisit serves as a practical complement to traditional metrics, offering empirical insight into which parts of the codebase developers find most difficult to understand.

\textbf{Potential for Integration with Software Development Tools:} NRevisit can be embedded into emergent AI-assisted code generation and code review systems. Using Tobii Eye Tracker 5L as a baseline, which provides a 133 Hz sampling rate and sub-degree precision, NRevisit enables real-time gaze tracking with minimal resource consumption. It operates efficiently with low CPU usage (5-10\%) and modest memory requirements (200-500MB), making it easy to set up and integrate without overwhelming system resources. In future adaptive software development environments, NRevisit can optimize real-time debugging, code navigation, and developer productivity, offering a lightweight, effective solution for enhancing software development workflows without significant computational overhead. 

\subsection{Limitations}
Despite the promising results of NRevisit in estimating cognitive load and code understandability, several limitations must be considered:

\textbf{Dependency on Eye-Tracking Accuracy:}
The effectiveness of NRevisit relies on the accuracy of eye-tracking devices. Variations in hardware quality, calibration errors and user movement can introduce noise into gaze data, potentially affecting revisit count calculations. While our study was conducted in a controlled environment, real-world usage in dynamic workspaces may introduce variability in data quality.

\textbf{Cognitive Load Measurement Validation:}
EEG is widely regarded as a reliable measure of cognitive load, but it has inherent limitations, including signal noise, individual variability and potential external influences (e.g., fatigue and emotional state). Although NRevisit demonstrated strong correlations with EEG-measured cognitive load, additional validation using alternative cognitive load assessment methods (e.g., task performance, subjective workload ratings) could further solidify our results.

\textbf{Generalizability Across Programming Tasks:}
This study focused on code comprehension tasks in a controlled setting, with a limited number of Java programs and structured code regions. While NRevisit was found to be highly predictive of cognitive load, its effectiveness across other programming paradigms (e.g., debugging, code writing, refactoring) remains to be validated. Future research should explore its applicability in diverse software development tasks.

\textbf{Expertise-Level Differences and Adaptability:}
The results indicate that expert and novice programmers exhibit different gaze revisit behaviors. While NRevisit successfully differentiated between expertise levels, further research is needed to determine how this metric can be adapted for personalized complexity assessment. 

\textbf{Integration with Other Complexity Metrics:}
Although NRevisit outperformed traditional complexity metrics in non-linear predictive models, a hybrid approach incorporating both static and behavioral metrics may provide an even more comprehensive code understandability assessment. Future work will explore the complementary of combining NRevisit with traditional metrics.

\subsection{Applications}
The findings from this study highlight several potential applications of NRevisit in software engineering and AI-assisted programming environments:

\textbf{AI-Assisted Code Review Systems:}
NRevisit can be integrated into automated code review tools to provide real-time insights into the cognitive effort required to understand code. This would allow reviewers and developers to prioritize complex sections of code that may require refactoring or additional documentation. AI-powered coding assistants could leverage gaze-based complexity assessment to dynamically recommend improvements.

\textbf{Adaptive Learning and Programming Training:}
NRevisit can be utilized in educational platforms to assess student engagement and learning difficulty when reading and understanding code. Adaptive learning systems could personalize exercises based on individual cognitive load, ensuring that programming challenges are neither too simple nor too complex for students.

\textbf{Code Complexity Estimation for AI-Generated Code:}
As AI-generated code becomes more prevalent, NRevisit can provide a behavioral assessment of AI-written code quality by evaluating its cognitive complexity for human developers.
This would be particularly beneficial for evaluating the understandability of AI-generated functions and improving AI-assisted code-generation tools.

\textbf{Real-Time Complexity Feedback in IDEs:}
Integrated Development Environments (IDEs) could incorporate NRevisit to provide real-time complexity feedback as developers navigate code. This would help programmers identify problematic code segments, reducing cognitive strain and improving code maintainability.

\textbf{Enhancing Cognitive Load Analysis in Human-Centered Software Engineering:}
NRevisit can be used in research focusing on human factors in software engineering, particularly in understanding cognitive effort in large-scale software projects. Future studies could integrate NRevisit with biometric measures such as heart rate variability or pupillometry for a multimodal approach to cognitive load assessment.

\section{Conclusion}
Understanding and measuring code understandability is a critical challenge in software engineering. Traditional static complexity metrics, while widely used, fail to capture the cognitive effort required for code comprehension. In this study, we introduced NRevisit, a behavioral complexity metric derived from eye-tracking revisit patterns, to dynamically assess cognitive load.

By correlating NRevisit with EEG-measured cognitive load, our results demonstrate that gaze revisits serve as a strong indicator of code complexity, outperforming traditional static metrics in predictive accuracy and robustness. Through comparative analysis, we established that the Close Loop definition of NRevisit (CL NRevisits) —which accounts for meaningful gaze transitions within code regions—provides a more reliable and interpretable representation of cognitive engagement than a continuous time-based revisit measure.

Our experiments further reveal that CL NRevisit significantly enhances cognitive load prediction when used in Neural Networks (NN) and Gaussian Process Regression (GPR) models, surpassing traditional complexity metrics in non-linear predictive settings. While Halstead Effort (HEff) was more frequently selected in feature selection experiments, CL NRevisit contributed more often to high-performing models ($R^2 \geq 0.8$), emphasizing its unique value in behavioral complexity assessment.

These findings suggest integrating NRevisit into software engineering workflows can provide a more human-centered approach to complexity evaluation. Potential applications include AI-assisted code reviews, adaptive programming education and real-time complexity feedback in IDEs. However, further validation is required across diverse programming tasks and future research should explore hybrid approaches that integrate NRevisit with traditional static metrics for a more comprehensive complexity assessment.

By bridging the gap between structural complexity analysis and human cognitive effort, NRevisit lays the groundwork for next-generation complexity assessment tools, enabling more effective software development practices in AI-driven and human-centered programming environments.

Future research should validate NRevisit across diverse programming tasks (e.g., debugging, refactoring) and examine how expertise levels influence revisit patterns to develop adaptive complexity models. Integrating NRevisit into AI-assisted programming tools and exploring multimodal cognitive load estimation with biometric data can further enhance its applicability. As a human-centered alternative to traditional complexity metrics, NRevisit bridges behavioral and static measures, offering promising advancements for both research and real-world software development.

\section*{Data Availability}
The datasets generated and analyzed during the current study are available in the \textcolor{blue}{\href{https://github.com/SoftwareRepository2025/NReivisit}{GitHub Repository}}.

\section*{Acknowledgements}
This work is partially financed through national funds by FCT - Fundação para a Ciência e a Tecnologia, I.P., in the framework of the Project UIDB/00326/2025 and UIDP/00326/2025.

\printbibliography

\end{document}